\documentclass[a4paper,american,superscriptaddress,  endfloats*]{revtex4}
\usepackage[T1]{fontenc}
\usepackage[latin1]{inputenc}
\usepackage{subfigure}
\usepackage{amsmath}
\usepackage{graphicx}
\usepackage{amssymb}

\makeatletter

\newcommand{\noun}[1]{\textsc{#1}}

\usepackage{mathptmx}
\usepackage[scaled=0.9]{helvet}
\usepackage{courier}

\usepackage{babel}
\makeatother
\begin{document}

\title{Freezing and folding behavior \\
in simple off-lattice heteropolymers}

\author{J. E. Magee}

\affiliation{Department of Chemical Engineering}

\author{J. Warwicker}

\affiliation{Department of Biomolecular Sciences}

\author{L. Lue}

\affiliation{Department of Chemical Engineering, \\
UMIST, PO Box 88, Manchester, M60 1QD, UK}

\date{\today}

\begin{abstract}
We have performed parallel tempering Monte Carlo simulations using
a simple continuum heteropolymer model for proteins. All ten heteropolymer
sequences which we have studied have shown first-order transitions
at low temperature to ordered states dominated by single chain conformations.
These results are in contrast with the theoretical predictions of
the random energy model for heteropolymers, from which we would expect
continuous transitions to glassy behavior at low temperatures.
\end{abstract}
\maketitle

\section{\label{sec:Introduction}Introduction}

Many biological heteropolymers (in particular proteins) can adopt
a single, sequence dependent conformation (the fold, or native state)
under physiological conditions. This specificity allows groups of
atoms within such a molecule to be placed accurately in space, allowing
the vast range of precise biochemical functions within living organisms.
Upon altering the environment around the protein (e.g.~pH, solvent
quality, temperature, etc.), the molecule can be made to lose this
single conformation (i.e.~unfold or denature), adopting a fluctuating,
disordered state. This \emph{folding transition} has been shown to
be both reversible and first-order for many proteins \cite{foldingreversible},
and is still poorly understood.

The modern theoretical viewpoint on the thermodynamics of protein
folding (recently reviewed by Pande et al.~\cite{PandeReview}) uses
concepts from the theory of spin glasses. This approach was introduced
by Bryngelson and Wolynes \cite{bryngelson}, who used the random-energy
model (REM) \cite{REM} to describe the low temperature behavior of
heteropolymers. This simple model assumes the energies of the possible
conformations of a heteropolymer to be independent and uncorrelated,
and predicts a second order transition (no latent heat) to a glassy
state dominated by a low energy conformation. The more complicated
theory of Shakhnovitch et al.~\cite{ShakhnovichGutinTheory} uses
another trick from the spin glass repertoire, the replica approach,
to arrive at the same conclusions. 

The low energy conformation occupied in the glassy state is not necessarily
the native state. Depending upon details of conditions, sequence and
sample history, the heteropolymer may well become {}``misfolded'';
that is, locked in to a non-native, low energy conformation. Indeed,
most randomly selected sequences should have several closely-spaced
conformations at the low end of their energy spectrum, and will not
reliably or reproducibly fold into only one of these. Some annealing
of the heteropolymer \emph{sequence} (i.e.~design work) should be
needed to {}``drive down'' the energy of a chosen (target) conformation
well below the closely-spaced region of the energy spectrum, preventing
sampling of competing conformations and allowing robust folding \cite{PandeReview}.
Simulations \cite{PERMlatticesim,PandeReview} and numerical studies
\cite{sglatticesim} using lattice heteropolymer models support these
predictions and the approximations behind the models. 

However, the random energy approximation is essentially mean field
in nature, and cannot accurately treat systems with persistent correlations
\cite{REMproblems}. Further, lattice models artificially constrain
the configurations available to a polymer. While lattice homopolymer
results have been instructive at high temperatures \cite{GrosbergText,FloryText},
they do not predict the low temperature freezing transition seen in
the simulations of Zhou et al.~\cite{hallpaper,exthallpaper}, where
an isolated homopolymer freezes into a single, apparently unique crystalline
state via a first-order transition. 

We have performed Monte Carlo simulations using ten random sequence
realizations of a simple off-lattice heteropolymer model to test these
predictions in continuum systems. We report that, rather than exhibiting
glassy behavior at low temperature, all ten sequences studied pass
through a first-order transition to an ordered low energy state, dominated
by a single conformation, without need for sequence design work.

\section{\label{sec:Model}Model}

\begin{figure}[p]
\includegraphics{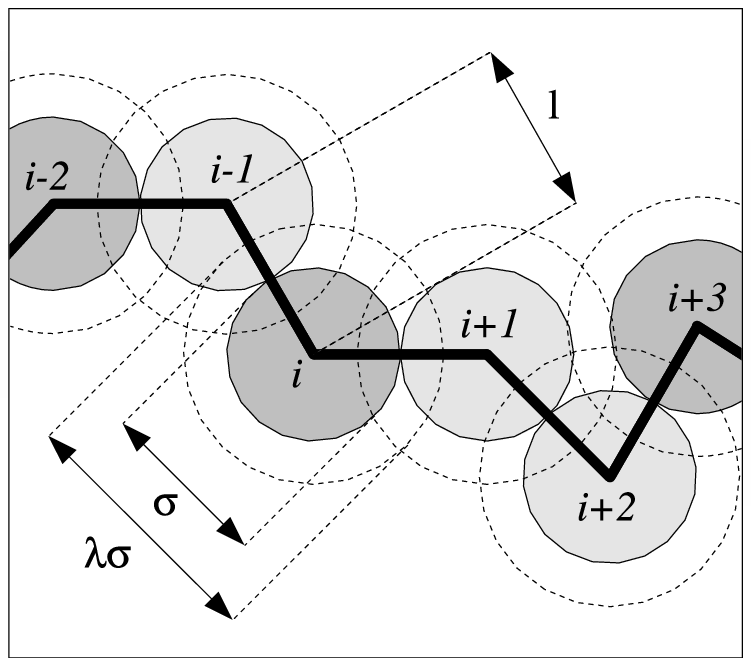}

\caption{\label{cap:MolFig} Schematic drawing of molecular model; a system
of spherical monomers (filled circles), diameter $\sigma$. Monomers
are joined together in a non-branching linear sequence with bonds
of length $\sigma(1-\delta)<l<\sigma(1+\delta)$. Non-bonded monomers
interact if they are within distance $\lambda\sigma$ of each other
(shown with dotted circles); in the figure, monomers $i+1$ and $i+2$
are interacting.}
\end{figure}

Our heteropolymers are modeled as freely-jointed chains of $N$ hard-sphere
monomers of diameter $\sigma$, each assigned a type; we denote the
type of monomer $i$ by $s_{i}$. The model is a simple extension
of that used by Zhou et al.~\cite{hallpaper,exthallpaper} and is
shown schematically in Fig.~\ref{cap:MolFig}.%
 Bonded monomers interact via the potential 

\begin{equation}
u_{i,i+1}(r)=\left\{ \begin{array}{ccccccc}
\infty &  &  &  & r & < & (1-\delta)\sigma\\
0 &  & (1-\delta)\sigma & < & r & < & (1+\delta)\sigma\\
\infty &  & (1+\delta)\sigma & < & r\end{array}\right.\label{eq:bondedpotn}\end{equation}

\noindent The parameter $\delta$ is a bond-length flexibility parameter;
this allows for conversion to molecular dynamics simulation and has
been shown to make no phenomenological difference to homopolymer folding
results \cite{hallpaper,exthallpaper}. We use $\delta=0.1$, allowing
a ten percent variation in bond length.

If monomers $i$ and $j$ are not directly bonded (that is, $\left|i-j\right|>1$),
then they interact via a square-well potential:

\begin{equation}
u_{ij}(r)=\left\{ \begin{array}{ccccccc}
\infty &  &  &  & r & < & \sigma\\
\epsilon_{s_{i}s_{j}} &  & \sigma & < & r & < & \lambda\sigma\\
0 &  & \lambda\sigma & < & r\end{array}\right.\label{eq:nonbondedpotn}\end{equation}

\noindent where $r$ denotes the distance between monomers $i$ and
$j$, $\lambda\sigma$ is the square well width, and $\epsilon_{ss'}$
is an \emph{interaction matrix} giving square-well interaction strengths
(square-well depths) between the monomer types. In this work, we use
a modified hydrophobic-polar, or \emph{mHP} matrix %
\footnote{An unmodified hydrophobic-polar matrix has polar-type monomers interacting
as hard spheres. This form for the interaction leads to P-type rich
sections of the heteropolymer sequence forming long loops off of a
central spherical nucleus rich in hydrophobic-type monomers \cite{iterativecolouring};
this behavior is not seen in real proteins.%
}. This is given by:

\begin{equation}
\epsilon^{mHP}=\epsilon'\left(\begin{array}{cc}
-1 & -\frac{1}{2}\\
-\frac{1}{2} & 0\end{array}\right)\label{eq:hpmatrix}\end{equation}

\noindent where $\epsilon'$ is the characteristic energy scale. With
this model, monomers with $s_{i}=0$ are hydrophobic-like (H-type),
preferring type-symmetric interactions, while type $s_{i}=1$ monomers
are polar-like (P-type), preferring type-asymmetric interactions.
Similar interaction matrices are widely used in on-lattice protein
folding simulations and numerical work \cite{interactionmatrices,latticeenumeration,dillestimate1}
and are believed to be good approximations to real intraprotein interactions
\cite{HPcollapse}. In the dense, globular heteropolymer state, we
expect this interaction matrix to lead to configurations with a core
of hydrophobic monomers surrounded by polar monomers, as seen in globular
proteins.

We use reduced units scaled by the hard-sphere radius $\sigma$ and
the energy scale $\epsilon'$, such that

\begin{equation}
\begin{array}{ccc}
r^{*} & = & r/\sigma\\
E^{*} & = & E/\epsilon'\\
T^{*} & = & k_{B}T/\epsilon'\end{array}\label{eq:reducedunits}\end{equation}

Polypeptides of length greater than approximately 30 amino acids can
form stable folded domains (e.g. the 35 amino acid WW domain \cite{wwdomain}).
We have chosen to study $N=64$ length heteropolymers; this length
is short enough for simulation in reasonable time, while long enough
that equivalent real proteins can fold. We have studied ten random
sequences interacting via the \emph{m}HP interaction matrix, each
with a $1:1$ H:P composition ratio. If we consider P-type monomers
to map onto charged and polar amino acid residues, and H-type monomers
to map to all other amino acid residues, this composition ratio is
similar to that found in real proteins \cite{composition}.

\section{\label{sec:Simulation-Method}Simulation Method}

Monte Carlo simulations were performed for a single polymer in isolation.
As such, we have not used periodic boundary conditions. We have implemented
crankshaft, pivot and continuum configurational bias (CCB) polymer
moves \cite{SadusBook}. Bond length fluctuation is allowed both in
the CCB move implementation, as well as through standard Monte Carlo
particle displacement moves performed on the monomers. CCB moves regrow
end segments of up to four monomers. Moves are performed with relative
probabilities in the ratio $N:(1/2):(1/4):(1/4)$ for particle displacement,
crankshaft, CCB and pivot moves respectively. Initial equilibration
runs last $3\times10^{9}$ Monte Carlo move attempts; data collection
runs last $1\times10^{9}$ move attempts.

The Monte Carlo simulations also use parallel tempering (also known
as replica exchange) \cite{SmitandFrenkel}. Several Monte Carlo simulations
of the same system at differing temperatures are run in parallel;
these simulations attempt to exchange configurations at regular intervals.
This helps to maintain the ergodicity of the simulations; an individual
system which has become stuck at low temperature in a metastable {}``trap''
may be promoted to higher temperatures, where it can explore a wider
range of phase space, aiding its equilibration. Our parallel tempering
setup uses ten individual simulations across the temperature range
$T^{*}=0.15\ldots1.5$, distributed such that the ratio between consecutive
temperatures is constant \cite{temperingdistn}. Configuration swap
moves are attempted every $1000$ Monte Carlo move attempts.

The simulations sample the configurational energy $E$, the radius
of gyration $r_{g}$, the two-particle density $\rho^{(2)}(r)$, and
the instantaneous contact map $\mathbf{c}$ for the system every $10^{5}$
Monte Carlo move attempts. The two-particle density is defined as 

\begin{equation}
\rho^{(2)}(r)=\frac{1}{N}\left\langle \sum_{i}\sum_{j\neq i}\delta\left(r\textnormal{-}r_{ij}\right)\right\rangle \label{eq:altg(r)}\end{equation}

\noindent where $r_{ij}$ is the distance between monomers $i$ and
$j$ \cite{HansenMcDonald}. 

The instantaneous contact map $\mathbf{c}$ reflects a heteropolymer
conformation; it is an $N\times N$ matrix, with each element $c_{ij}$
equal to one if monomers $i$ and $j$ are within each others square
well, and zero otherwise. The instantaneous contact map is therefore
given by:

\begin{equation}
c_{ij}=\left\{ \begin{array}{ccc}
0 &  & r_{ij}<\lambda\sigma\\
1 &  & r_{ij}>\lambda\sigma\end{array}\right.\label{eq:contactmap}\end{equation}

The average contact map $\left\langle c_{ij}\right\rangle $ gives
us information on the structure adopted by the system at given temperature.
We can use it to calculate a \emph{normalized thermal average overlap}:

\begin{equation}
\left\langle \mathcal{Q}_{n}\right\rangle =\frac{{\displaystyle \sum_{i\neq j}\left\langle c_{ij}\right\rangle ^{2}}}{{\displaystyle \sum_{i\neq j}\left\langle c_{ij}\right\rangle }}\label{eq:overlap}\end{equation}

\noindent This normalized overlap parameter can take values between
zero and one. Values close to one indicate that the system is dominated
by only a very small number of very similar conformations; small values
indicate that the system is exploring a variety of different conformations.

Energy and radius of gyration data from the individual simulations
within a parallel tempering run are tied together using self consistent
multiple histogram extrapolation \cite{multihistogram}. This allows
the histograms of energy $E$ and other observables $\{ A\}$ recorded
from the individual simulations to be combined to estimate $\Omega(E,\{ A\})$,
the density of states of the system. From this, we can interpolate
the behavior of the system across, and to a certain extent beyond,
the simulated temperature and parameter range. We use the extrapolated
distribution at given temperature to calculate the constant volume
specific heat $c_{v}^{*}$, given by

\begin{equation}
c_{v}^{*}=\frac{\left\langle E^{*2}\right\rangle -\left\langle E^{*}\right\rangle ^{2}}{T^{*2}}\label{eq:reducedcv}\end{equation}

Histogram extrapolation methods rely on good sampling across the range
of interest by the individual simulations. As such, simulations around
first-order transitions with a significant free energy barrier between
phases can be problematic. If simulations on either side of the transition
temperature sample only the phase which is most stable at that temperature,
but no simulation actually observes the system passing back and forth
between phases, the extrapolation process can {}``blur out'' the
transition, leaving a signature only in an increase in the error estimate
and the response functions (e.g.~specific heat) across the region
of the transition. Worse still, if the system exhibits strong metastability,
this can introduce a large systematic error into any estimate for
the position of the transition. 

To avoid this problem, we have used multicanonical sampling \cite{multicanonical}
to directly confirm or deny phase coexistence where we suspect the
existence of a first-order transition. We have used an \emph{iterative
multicanonical scheme} \cite{MulticanonicalSchemes}. Repeated biased
Monte Carlo simulations refine our estimate for the density of states
of the system, with the aim of sampling evenly in energy across the
suspected range of coexistence. The bias can then be removed from
the results of these simulations, and the refined estimate for the
density of states can be spliced back into the wider simulation results.
The individual simulations used in this procedure are performed using
the data collection run parameters stated above.

\section{\label{sec:Simulation-Results}Simulation Results}

\begin{figure}
\includegraphics[%
  scale=0.4]{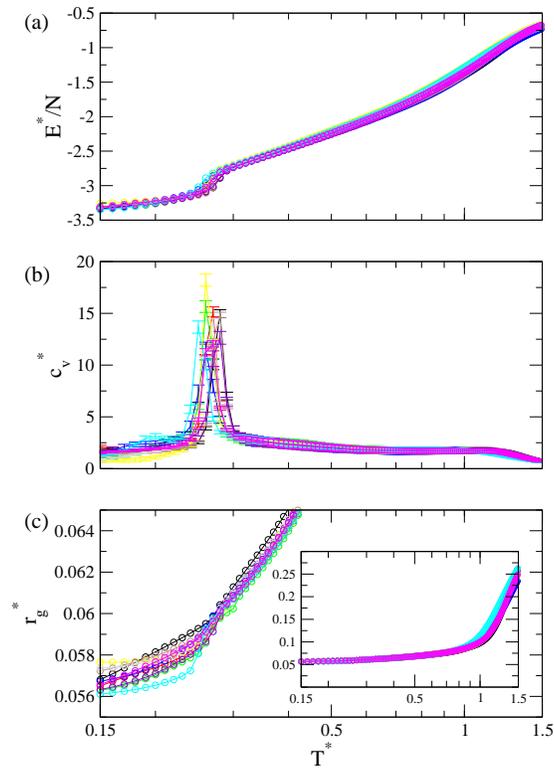}

\caption{\label{cap:HPSeqDat} \noun{}(a) Energy per unit monomer, (b) heat
capacity and (c) radius of gyration as a function of temperature for
the ten sequences studied. The scale of plot (c) has been chosen to
show the low temperature behavior; the inset shows the $r_{g}$ behavior
over the full temperature range studied. Curves are calculated at
$T^{*}$ intervals of $0.01$ using $\Omega\left(E,r_{g}\right)$
calculated from parallel tempering runs as described in the text.
Estimated errors are smaller than symbol size for radius of gyration
and energy data.}
\end{figure}

\begin{figure}
\includegraphics[%
  scale=0.5]{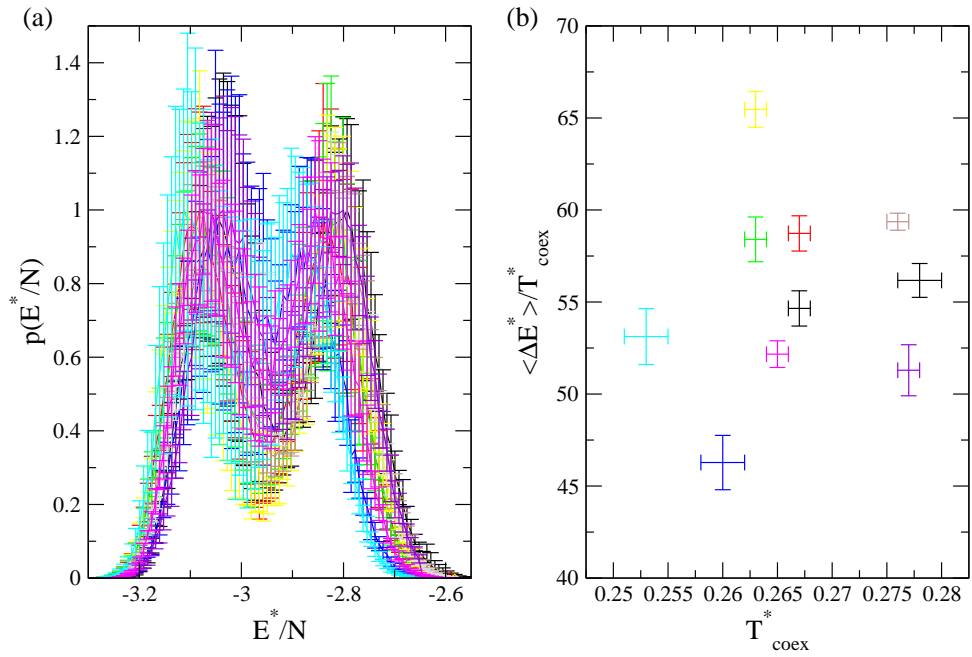}

\caption{\label{cap:HPseqcoex} (a) Coexistence energy distributions $p\left(E^{*}/N\right)$for
the sequences studied. (b) Coexistence temperature $T_{coex}^{*}$
plotted against energy difference $\Delta E/k_{B}T$ between the coexisting
phases. Color denotes sequence in both graphs. \noun{}}
\end{figure}

Plots of $E^{*}$, $c_{v}^{*}$ and $r_{g}^{*}$ for our sequences,
interpolated across the studied temperature range, are shown in Fig.~\ref{cap:HPSeqDat}.%
 The phenomenology does not vary considerably with sequence. Above
a temperature of $T^{*}\approx0.9$, the heteropolymer exists in a
disordered coil state - a high energy structure that rapidly increases
its radius of gyration and decreases its heat capacity (energy fluctuations)
with temperature. Between $T^{*}\approx0.3$ and $T^{*}\approx0.9$,
we see the expected low temperature collapsed globule state with a
small radius of gyration (order of $0.05\sigma$) and with relatively
high heat capacity (order of $2.0k_{B}$). Around $T^{*}\approx0.26$,
however, we see sharp spikes in the specific heat, associated with
sudden drops in both conformational energy and radius of gyration
with decreasing temperature, suggestive of first-order transitions.
We have used multicanonical sampling to study these apparent transitions;
for all sequences, we have found temperatures at which double peaked
probability distributions occur for conformational energy, indicating
two-phase coexistence. These probability distributions are shown in
Fig.~\ref{cap:HPseqcoex}(a).%
 The sequence dependence of the probability distributions is small.

In Fig.~\ref{cap:HPseqcoex} we plot coexistence temperatures for
the sequences studied against the energy difference $\Delta E^{*}/T^{*}$
between the coexisting phases. The quantities are clustered around
$T^{*}\approx0.27$ and $\Delta E/k_{B}T\approx56$, but otherwise
appear uncorrelated. For comparison, the typical latent heat $\Delta H$
of the folding transition for a globular protein is of order of $100k_{B}T$
\cite{foldingheat}.

\subsection{Structural character of the low energy phase}

\begin{figure}
\includegraphics[%
  scale=0.5]{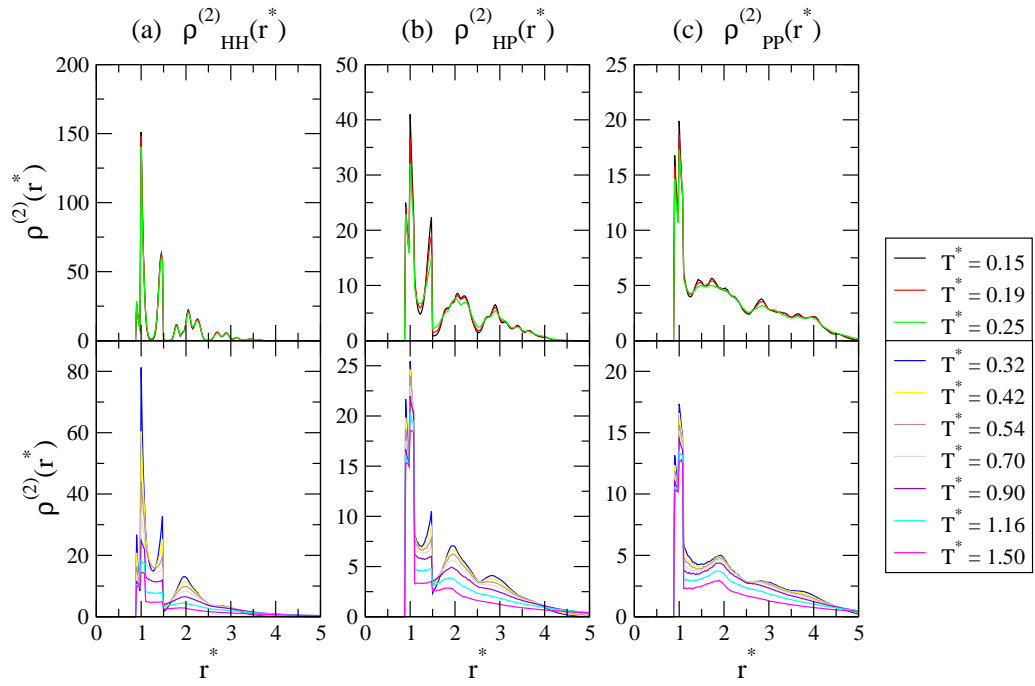}

\caption{\label{cap:seq0g(r)} Two-particle densities for sequence 4 ($T_{coex}^{*}=0.263$).
In each case, the top graph shows $\rho^{(2)}\left(r^{*}\right)$
for temperatures below the observed transition, and the bottom graph
shows $\rho^{(2)}\left(r^{*}\right)$ for temperatures above the observed
transition. Discontinuities are present in the data at $r^{*}=0.9$
and $r^{*}=1.1$ due to the discontinuities in the direct binding
potential Eq.~(\ref{eq:bondedpotn}), and at $r^{*}=1.0$ and $r^{*}=1.5$
due to the discontinuities in the square-well potential Eq.~(\ref{eq:nonbondedpotn}).
Curves are color coded according to the temperature at which the data
was collected, as shown in the legend: (a) $\rho_{HH}^{(2)}\left(r^{*}\right)$,
(b) $\rho_{HP}^{(2)}\left(r^{*}\right)$, (c) $\rho_{PP}^{(2)}\left(r^{*}\right)$. }
\end{figure}

\begin{figure}
\subfigure[]{\includegraphics{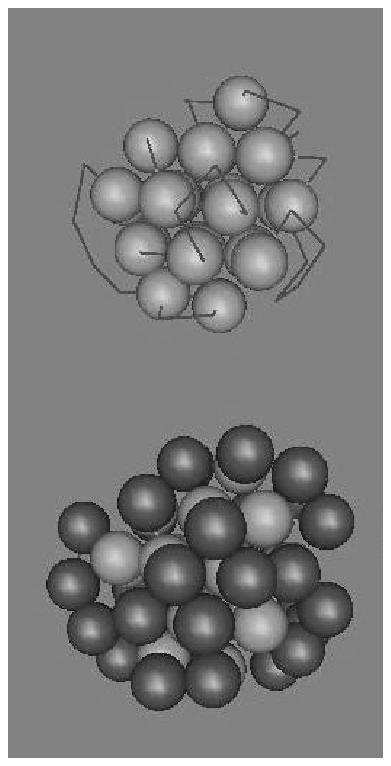}}\subfigure[]{\includegraphics{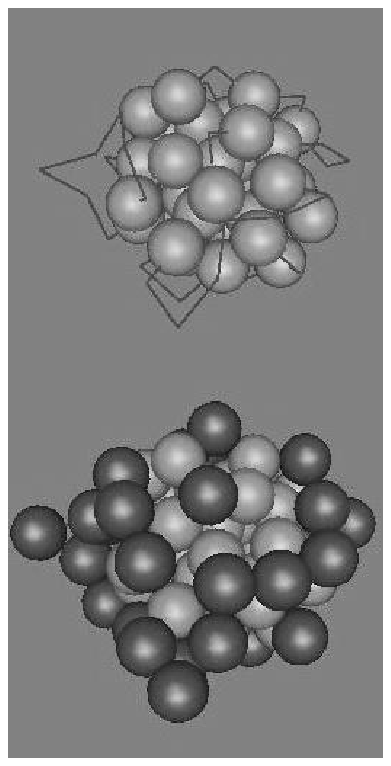}}

\caption{\label{cap:hexplaneconfig} Typical heteropolymer configurations,
generated with gOpenMol \cite{gOpenMol1,gOpenMol2}. Pale spheres
represent hydrophobic monomers, dark spheres represent polar monomers.
(a) Typical low energy phase configuration, with perspective chosen
to clearly show hexagonal organization of hydrophobic monomers. In
the top picture, display of polar monomers has been suppressed to
better show the hydrophobic core. (b) Typical disordered globular
phase configuration. In the top picture, display of polar monomers
has been suppressed to better show the hydrophobic core.}
\end{figure}

We present two-particle densities $\rho^{(2)}$ data, calculated using
Eq.~(\ref{eq:altg(r)}), for sequence 4 in Fig.~\ref{cap:seq0g(r)};%
 these plots are typical of the results for all ten sequences studied.
These data were collected before the multicanonical algorithm was
used, and may exhibit some degree of metastability. We first consider
$\rho_{HH}^{(2)}\left(r^{*}\right),$the hydrophobic-hydrophobic two-particle
densities, in Fig.~\ref{cap:seq0g(r)}(a). We see that below the
transition temperature, the first three neighbor shells are clear
and distinct, separated by regions with near-zero hydrophobic particle
density. The next three neighbor shells are much more sparsely populated
(limited by the diameter of the collapsed state), but remain distinct,
separated by sharp minima. These data show a low energy phase structure
with local translational order between hydrophobic monomers. For $\rho_{HH}^{(2)}\left(r^{*}\right)$
above the transition, the first three neighbor shells are still clear
up to $T^{*}=0.7$, but not distinctly separated. The next three neighbor
shells have blurred together into one single peak. Above the transition
temperature, the hydrophobic monomers have no translational order,
as expected in a disordered globular phase. We show typical low-energy
phase and globular phase configurations in Fig. \ref{cap:hexplaneconfig};%
 the orientation of the low-energy configuration has been chosen to
clearly show the hydrophobic monomers arranged in hexagonal crystal
planes, whereas no crystal structure exists in the globular configuration.

Figures \ref{cap:seq0g(r)}(b) and (c) show hydrophobic-polar and
polar-polar two-particle densities $\rho_{HP}^{(2)}\left(r^{*}\right)$
and $\rho_{PP}^{(2)}\left(r^{*}\right)$, respectively. Neither show
any sign of translational order, either above or below the transition.
The polar monomers remain in a disordered state across the temperature
range studied.

\begin{figure}
\includegraphics[%
  clip,
  scale=0.8]{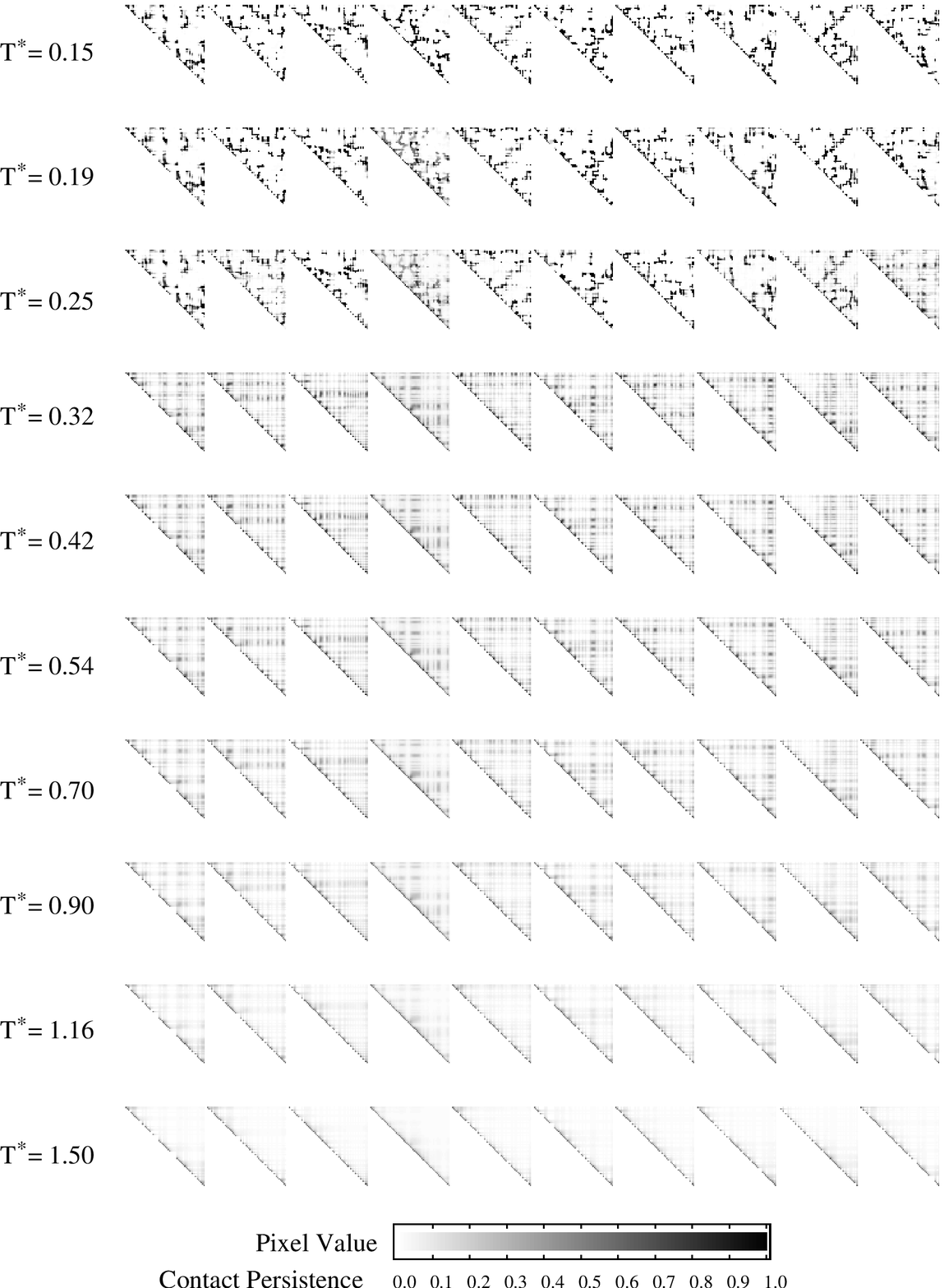}

\caption{\label{cap:cmaps} }

\noindent Figure \noun{\ref{cap:cmaps}:} Average contact maps $\left\langle \mathbf{c}\left(T^{*}\right)\right\rangle $
for the ten sequences studied. Each contact map is shown as a $64\times64$
pixel block, pixels corresponding to matrix elements, with the contact
persistence $\left\langle c_{ij}\right\rangle $ given by the shade
of the pixel as shown in the key. Contact maps are symmetric across
the diagonal $\left(\mathbf{c}=\mathbf{c}^{T}\right)$, so only the
upper half is shown. Contact maps in the same column are from the
same sequence. Contact maps in the same row were collected at the
same temperature. 
\end{figure}

We show the thermal average contact maps in Fig.~\ref{cap:cmaps}.%
 Note that these data are collected from equilibrated parallel tempering
runs before the multicanonical algorithm is used, and as such exhibit
metastability. At low temperatures (below $T^{*}\approx0.32$), we
see that the contact maps are dominated by very persistent contacts,
suggesting that the low temperature structure is dominated by very
few configurations. Contact maps collected at higher temperatures
show large low persistence regions, as expected for a disordered globular
state. 

\begin{figure}
\includegraphics[%
  scale=0.5]{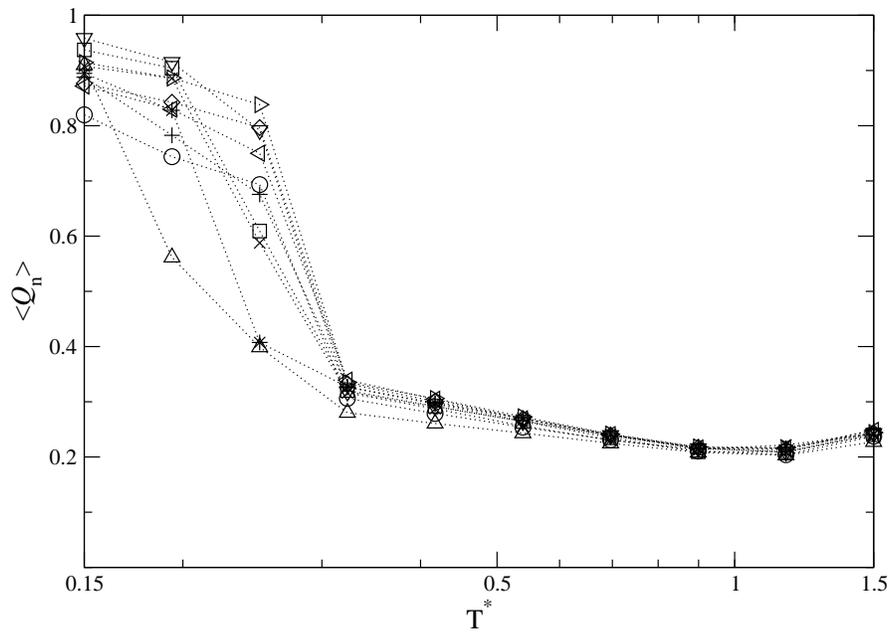}

\caption{\label{cap:overlap} Normalized thermal average overlap parameter
$\left\langle \mathcal{Q}_{n}\right\rangle $ for the ten mHP sequences
studied, calculated from the thermal average contact maps shown in
Fig \ref{cap:cmaps}. Different symbols denote different sequences.
Lines serve only as a guide to the eye.}
\end{figure}

To quantify this, plots of the normalized overlap parameter, calculated
from these contact maps, are shown in Fig.~\ref{cap:overlap}.%
 As with the contact maps, these results exhibit metastability around
the transition. For all sequences, we see a large increase on cooling
in $\left\langle \mathcal{Q}_{n}\right\rangle $ from values of $\left\langle \mathcal{Q}_{n}\right\rangle \approx0.3$
in the disordered globule phase to values of $\left\langle \mathcal{Q}_{n}\right\rangle \approx0.85$
in the low energy phase. This confirms that the low energy phase is
dominated by small fluctuations around a single conformation.

\section{\label{sec:Discussion-and-Conclusions}Discussion and Conclusions}

Our results indicate that, for all of the ten random heteropolymer
sequences which we have studied, there exists a low temperature transition
between the familiar collapsed liquid-like globule phase and a low
energy phase. These transitions do not have the glassy character expected
from replica theory, showing instead double-peaked energy distribution
functions with substantial associated latent heat. In each case, the
low energy phase is more compact than the disordered globule, having
a lower $r_{g}$, exhibits local translational order and crystal planes,
as shown by the two-particle density function $\rho^{(2)}$ (similar
to the low energy phases seen in off-lattice homopolymer simulations
\cite{hallpaper,exthallpaper}), and is not glassy, being dominated
by a single conformation of the polymer backbone, as shown by the
normalized thermal average overlap parameter $\left\langle \mathcal{Q}_{n}\right\rangle $.
In analogy to the more familiar phase behavior of bulk systems, we
describe the transition as a freezing transition, and the low-energy
state as a frozen state. We note the obvious parallels with protein
folding transitions, where polypeptide heteropolymers undergo a first-order
phase transition to a collapsed single conformation, exhibiting repeating
structural units.

The large computational effort associated with simulating high-density,
low temperature continuum polymer states has limited the number of
sequences we have been able to study. Nevertheless, we can use the
Newcombe-Wilson score method (method 3 in Ref. \cite{newconfderiv})
to estimate with 95\% confidence that \emph{at least} 72.25\% of 64
length 1:1 HP composition ratio sequences will demonstrate an equivalent
first-order freezing transition to a non-glassy single conformation
state. 

We suggest that the reason that we see different behavior from equivalent
lattice models, which undergo low temperature glass transitions, is
a matter of symmetries. As evidenced by the $\rho^{(2)}$ data (see
Fig.~\ref{cap:seq0g(r)}), the disordered globular state for our
sequences has no translational order, with hydrophobic monomers able
to occupy any position, within the bounds of bond lengths and hard-sphere
overlap. Passing below the freezing temperature, the frozen states
of our sequences gain crystalline-like order %
\footnote{We say crystalline-like rather than crystalline since it is possible
that, as in case of two dimensional crystals, this translational order
is not long-ranged, and in the large $N$ limit would be destroyed
by fluctuations. However, this will be unimportant for polymers which
freeze into objects with short length scales, which we note includes
real proteins as well as the simple model heteropolymers studied here. %
}; hydrophobic monomers must occupy lattice points, or points very
near them. The spatial symmetry of the system has been broken in a
manner which, in three dimensions, requires a first-order phase transition.
In the lattice model, the approximation is that spatial symmetry is
already broken for the system at all temperatures, and there is no
possibility of such an order-disorder transition. Further, the symmetry
of the lattice is fixed in advance; if the frozen state of the off-lattice
system is on a different lattice, or if some distortion of the lattice
is necessary for stability, the lattice model cannot accurately portray
the ordered frozen state. With cooling, the lattice system will remain
in a disordered state, eventually forming a glass. 

The crystalline nature of the frozen states seen here do not resemble
the ordered structures seen in biological heteropolymers (i.e., proteins);
we see no signature of large scale helices or $\beta$-sheets in the
contact maps. We believe this to be due to the complete flexibility
of the chains in the model. It has been suggested \cite{optimalhelix}
that helices and $\beta$-sheet like {}``saddle'' structures are
optimal shapes for a stiff stringlike object, in the same manner that
the ubiquitous face centered cubic crystal structure is the optimal
arrangement for sets of spheres. We are currently investigating this
matter for both homo- and heteropolymeric systems. 

It can be seen from the contact maps (see Fig.~\ref{cap:cmaps})
that the frozen configurations are strongly sequence dependent. However,
the thermodynamic behavior of the heteropolymers (Fig.~\ref{cap:HPSeqDat})
has very little (though still statistically significant) sequence
dependence. We believe that the thermodynamic behavior, in the HP
model at least, has stronger composition dependence than sequence
dependence. Our 1:1 H:P composition ratio (with its freezing temperatures
bunched around $T^{*}\approx0.27$) lies between the limits of the
pure hydrophobic homopolymer (with a freezing temperature of $T^{*}=0.33$
\cite{hallpaper}), and the pure polar homopolymer with its effectively
hard-sphere monomers, which should not freeze in isolation. 

It is generally suggested that the fraction of heteropolymeric sequences
capable of freezing to unique conformations should be small, around
$10^{-9}$ \cite{molbio}. Dill and coworkers estimate the chance
of finding a protein sequence which will fold to a given approximate
structure as being of order $10^{-10}$ \cite{dillestimate1,dillestimate2,dillreview}.
Recent experimental work, using a large random-sequence protein library
to find sequences which bind to ATP, estimates the fraction of protein
sequences which are {}``functional'' as around $10^{-11}$ \cite{functionestimate}.
For our system, we have estimated that over 70\% of 64-length 1:1
composition ratio HP sequences freeze to unique conformations. Similar
over-estimation is also seen for lattice models; the impressive direct
enumeration performed by Li et al.~\cite{latticeenumeration} on
an HP lattice model suggests that as many as 4.75\% of all 27-length
HP sequences have unique ground state configurations. 

We believe such over-estimation to be due to a lack of sources of
frustration in our model, as compared to real protein systems. Our
model has only two monomer types, no size polydispersity or side chains,
completely flexible bonds, and no orientational dependence in its
interaction potential. Any of these factors could significantly lower
the number of model sequences which exhibit folding-like behavior.

In summary, we have demonstrated that a simple continuum heteropolymer
model can exhibit a first-order transition to an ordered low temperature
state dominated by a single conformation, without glassy behavior.
This behavior is similar to that seen in protein folding. It would
be of interest to see whether extensions to this model yield further
protein-like character, for example secondary structure, or instead
introduce frustrations which prevent the observed behaviors.

\begin{acknowledgments}
This work is supported by the BBSRC (grant reference B17005).
\end{acknowledgments}
\bibliographystyle{apsrev}
\bibliography{refs}

\end{document}